\documentclass[article,shortnames,nojss]{jss}\usepackage[]{graphicx}\usepackage[]{color}
\makeatletter
\def\maxwidth{ %
  \ifdim\Gin@nat@width>\linewidth
    \linewidth
  \else
    \Gin@nat@width
  \fi
}
\makeatother

\usepackage{amsmath,amssymb,mathrsfs}
\usepackage{subcaption}
\usepackage[T1]{fontenc}
\usepackage[utf8]{inputenc}

\newcommand{\ie}{{\itshape i.e.}}

\newcommand{\bs}[1]{\boldsymbol{#1}}
\newcommand{\CP}{C\!P}  
\newcommand{\card}[1]{n\left(#1\right)} 

\usepackage{alltt}
\renewenvironment{Sinput}{\begin{alltt}}{\end{alltt}}

\author{Peter E.\ DeWitt\\University of Colorado\\ Anschutz Medical Campus \And
        Samantha MaWhinney\\University of Colorado\\ Anschutz Medical Campus \And
        Nichole E.\ Carlson\\University of Colorado\\ Anschutz Medical Campus}
\title{\pkg{cpr}: An \proglang{R} Package For Finding Parsimonious B-Spline
Regression Models via Control Polygon Reduction and Control Net Reduction}

\Plainauthor{Peter E.\ DeWitt, Samantha MaWhinney, Nichole E.\ Carlson}
\Plaintitle{cpr: An R Package For Finding Parsimonious B-Spline Regression Models via Control Polygon Reduction and Control Net Reduction}
\Shorttitle{\pkg{cpr}: Control Polygon Reduction}

\Abstract{
  The \proglang{R} package \pkg{cpr} provides tools for selection of
  parsimonious B-spline regression models via algorithms coined `control polygon
  reduction' (CPR) and `control net reduction' (CNR).
  B-Splines are commonly used in regression models to smooth data and
  approximate unknown functional forms.  B-Splines are defined by a polynomial
  order and a knot sequence.  Defining the knot sequence is non-trivial, but is
  critical with respect to the quality of the regression models.  The focus of
  the CPR and CNR algorithms is to reduce a large knot sequence down to a
  parsimonious collection of elements while maintaining a high quality of fit.
  The algorithms are quick to implement and are flexible enough to support many
  types of data and regression approaches.  The \pkg{cpr} package provides the end user
  collections of tools for the construction of B-spline basis matrices,
  construction of control polygons and control nets, and the use of diagnostics
  of the CPR and CNR algorithms.
}


\Address{
  Peter E.\ DeWitt\\
  Department of Biostatistics and Informatics\\
  University of Colorado Anschutz Medical Campus\\
  Aurora Colorado, United States of America\\
  E-mail: \email{dewittpe@gmail.com} \\

  Samantha MaWhinney\\
  Department of Biostatistics and Informatics\\
  University of Colorado Anschutz Medical Campus\\
  Aurora Colorado, United States of America\\
  E-mail: \email{sam.mawhinney@ucdenver.edu} \\

  Nichole Carlson\\
  Department of Biostatistics and Informatics\\
  University of Colorado Anschutz Medical Campus\\
  Aurora Colorado, United States of America\\
  E-mail: \email{nichole.carlson@ucdenver.edu} \\
}

\IfFileExists{upquote.sty}{\usepackage{upquote}}{}
\begin{document}

\section{Introduction} \label{sec:intro}
Since their formal definition by \citet{curry1947}, B-splines have become
a common tool for approximating functions and surfaces with piece-wise
polynomials.  The fields of numeric analysis, computer aided graphics and
design, and statistics, to name only a few, have all benefited from B-splines.
The richness of the B-spline literature is
partially due to the 
lack of an analytic solution for optimal specification of the knot sequence
defining the B-splines and the resulting spline functions
\cite{jupp1978approximation}.

Methods for knot sequence specification vary based on secondary objectives.
Controlling the wiggliness of the spline function has been done by using knot
sequences of large cardinality and maximizing a penalized objective function
(see \citet{osullivan1986a} and penalized B-splines of \citet{eilers1996}, and
\citet{eilers2010}).
Estimation of the target function via the
(regression) coefficients instead of the spline function results in a larger
knot sequences \citep{lyche1988b}.
If we allow the location of the knots to be free then there are many different
adaptive methods \citep{biller2000, bakin1997, friedman1995, miyata2003adaptive,
ruppert2000, richardson2008multivariate, zhou2001spatially} for
knot sequences specification.

What we found to be lacking from the literature was a method for knot sequences
specification which was able to provide regression models with a high quality of
fit with a small number of degrees of freedom.  Further, we desired a model
selection approach that was computationally efficient for B-spline models of
uni-variable functions and extendible to multi-variable functions via tensor
products of B-splines.  For selection of such models, we developed the Control
Polygon Reduction (CPR) \citep{dewittp2017dissertation}, a backward-step
selection process based on the geometry of the control polygons about B-splines.
Computational efficiency is gained by operating on sparse low-rank matrices
instead of a dense high-rank design matrix.

\pkg{cpr}\footnote{Released Version:
  \url{https://cran.r-project.org/package=cpr}; Developmental Version:
\url{https://github.com/dewittpe/cpr}} is an \proglang{R}~\citep{R-base} package
developed to provided extended functions for B-splines, tensor products, control
polygons, control nets, and the model selection algorithms CPR and CNR.

This manuscript will focus on CPR and is structured as follows: Section~\ref{sec:background} provides a
brief overview of B-splines, control polygons, and the assessment of the influence
of a knot on a spline function.
The CPR algorithm is defined in Section~\ref{sec:cpr_algorithm}.
Section~\ref{sec:cpr_package} has detailed examples of the use of the functions
provided by \pkg{cpr} along with explanations for modeling functions of one
continuous predictor.  A brief overview of the CNR algorithm as associated tools
is provided in Section~\ref{sec:cnr}.

\section{Background} \label{sec:background}

Consider the following general model
\begin{equation}
  \label{eq:cpr_general_model}
  \bs{y} = f \left( \bs{x} \right) + \bs{Z}_{1} \bs{\beta} + \bs{Z}_{2} \bs{b} +
  \bs{\epsilon},
\end{equation}
where $\bs{y}$ is a $n \times 1$ vector of responses and
$\bs{x},$ another $n \times 1$
vector, is a continuous predictor.  The $\bs{Z}_{1}\bs{\beta}$ denotes
the $n \times p$ design matrix and $p \times 1$ coefficients vector for
(optional) fixed effects, $\bs{Z}_{2} \bs{b}$ the design matrix and coefficients for
(optional) random effects, and $\bs{\epsilon}$ the model, or measurement, error.
The function $f$ is the primary focus of our work, we aim to model this function
with parsimonious B-splines.

The control polygon reduction (CPR) model selection approach is a backward-step
selection process.  By starting with a large number of knots, CPR omits the
least influential knot at each step, between regression fits.

In this section we present an overview of B-splines, control polygons, and
introduce our metric for assessing the relative influence of an internal knot.
Additional detail on B-splines can be found in \citet{deboor2001} and
\citet{prautzsch2002}.

\subsection{Uni-variable B-splines and Control Polygons}
A B-spline basis matrix is defined by a polynomial order $k$ and knot sequence
$\bs{\xi}$ with the common construction of $k$-fold knots on the boundaries, set
to the minimum and maximum of the support, $l \geq 0$ interior knots, and sorted
in a non-decreasing order.  The matrix,
\begin{equation}
  \label{eq:basis_matrix}
  \bs{B}_{k, \bs{\xi}}\left(\bs{x}\right) =
  \begin{pmatrix}
    B_{1, k, \bs{\xi}} \left(x_1\right) & B_{2, k, \bs{\xi}} \left(x_1\right) & \cdots & B_{k + l, k, \bs{\xi}} \left( x_1 \right)  \\
    B_{1, k, \bs{\xi}} \left(x_2\right) & B_{2, k, \bs{\xi}} \left(x_2\right) & \cdots & B_{k + l, k, \bs{\xi}} \left( x_2 \right)  \\
    \vdots & \vdots & \ddots & \vdots \\
    B_{1, k, \bs{\xi}} \left(x_1\right) & B_{2, k, \bs{\xi}} \left(x_n\right) & \cdots & B_{k + l, k, \bs{\xi}} \left( x_n \right)
  \end{pmatrix},
\end{equation}
is constructed via de Boor's recursive formula:
\begin{equation}
  \label{eq:recurrence_relation}
  B_{j, k, \bs{\xi}}\left(x \right) =
  \omega_{j, k, \bs{\xi}}\left(x\right) B_{j, k-1, \bs{\xi}} \left(x \right) +
  \left(1 - \omega_{j+1, k, \bs{\xi}} \right) B_{j+1, k-1, \bs{\xi}}\left( x\right),
\end{equation}
with
\begin{equation}
  \label{eq:recurrence_relation_null}
  B_{j, k, \bs{\xi}}\left(x\right) = 0 \quad \text{for } x \notin
  \left[\xi_{j}, \xi_{j+k}\right), \quad
  B_{j, 1, \bs{\xi}}\left(x\right) = 
  \begin{cases}
    1 & x \in \left[\xi_{j}, \xi_{j+1} \right)\\
    0 & \text{otherwise},
  \end{cases} 
\end{equation}
and
\begin{equation}
  \label{eq:omega}
  \omega_{j, k, \bs{\xi}}\left(x\right) =
  \begin{cases}
  \hfil 0                                        & \hfil x \leq \xi_{j} \\
  \frac{x - \xi_{j}} {\xi_{j + k - 1} - \xi_{j}} & \xi_{j} < x < \xi_{j+ k - 1} \\
  \hfil 1                                        & \hfil \xi_{j + k - 1} \leq x
  \end{cases}.
\end{equation}

The basis matrix $\bs{B}_{k, \bs{\xi}}\left(\bs{x}\right)$ provides a partition
of unity over the support of $\bs{x},$
\ie, all rows sum to one,  and defines a spline space
$\mathbb{S}_{k, \bs{\xi}} = \text{span } \bs{B}_{k, \bs{\xi}}.$
The spline function, $f \left(x \right) \in \mathbb{S}_{k,
\bs{\xi}},$ is a convex sum the coefficients $\bs{\theta}_{\bs{\xi}},$
\begin{equation}
  \label{eq:bspline_univariable_function}
  f\left(x \right) = \bs{B}_{k,
  \bs{\xi}} \left( x \right) \bs{\theta}_{\bs{\xi}} =
  \sum_{j = 1}^{k + l} B_{j, k, \bs{\xi}} \left( x \right) \theta_{j, \bs{\xi}}.
\end{equation}

\begin{figure}
  \centering
  \begin{subfigure}[t]{0.48\textwidth}
\begin{Schunk}

\includegraphics[width=\maxwidth]{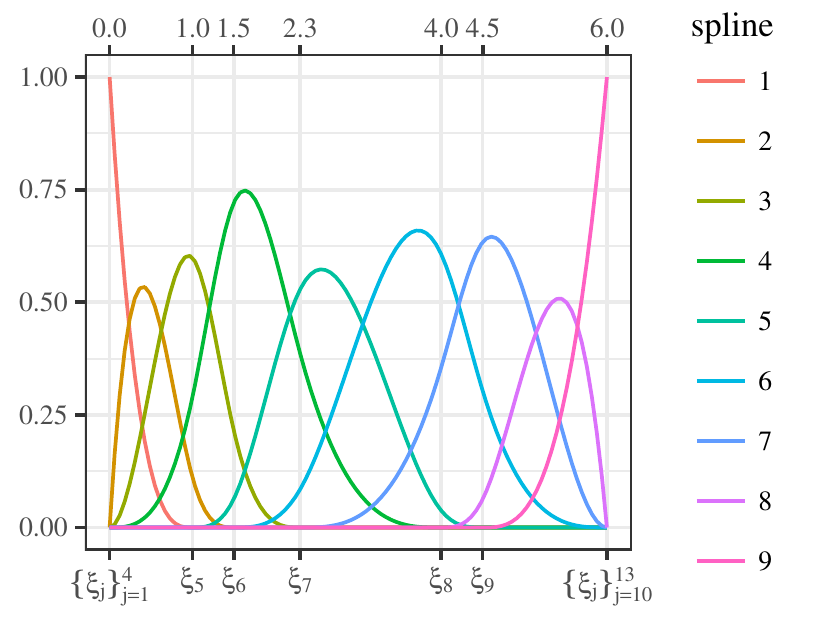} \end{Schunk}
  \caption{}\label{fig:basis_plot}
  \end{subfigure}
  \begin{subfigure}[t]{0.48\textwidth}
\begin{Schunk}

\includegraphics[width=\maxwidth]{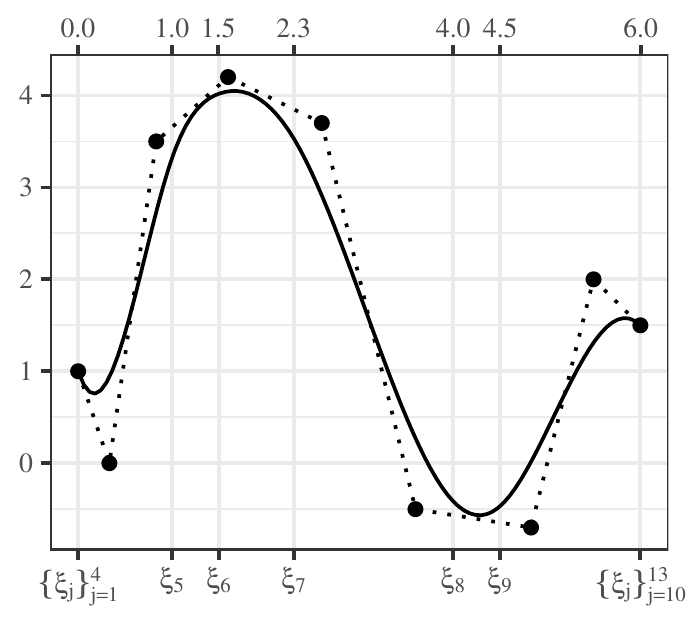} \end{Schunk}
  \caption{}\label{fig:cp_plot}
  \end{subfigure}
  \caption{(a) A B-spline basis, $\bs{B}_{k, \bs{\xi}} \left( x \right).$
  (b) A spline function $\bs{B}_{k, \bs{\xi}} \left( x \right)
  \bs{\theta}_{\bs{\xi}}$ (solid line), shown within its control polygon
  $\CP_{k, \bs{\xi}, \bs{\theta}_{\bs{\xi}}}$ (dotted line). Both (a) and (b) have the same
  basis, $k = 4$ order spline with knot sequence
  $\bs{\xi}$ = \{0, 0, 0, 0, 1, 1.5, 2.3, 4, 4.5, 6, 6, 6, 6\}. 
  The abscissae for the control vertices are $\bs{\xi}^{*}$ =
  \{0.00, 0.33, 0.83, 1.60, 2.60, 3.60, 4.83, 5.50, 6.00\} and the
  ordinates are $\bs{\theta}^{T}_{\bs{\xi}}$ = (1, 0, 3.5, 4.2, 3.7, -0.5, -0.7, 2, 1.5).}
  \label{fig:basis_and_cp_plot}
\end{figure}

There is no one-to-one association between the elements of the knot sequence,
cardinality $\card{\bs{\xi}} = 2k + l$ and the coefficients,
$\card{\bs{\theta}_{\bs{\xi}}} = k + l.$  However, a meaningful geometric
relationship between $\bs{\xi}$ and $\bs{\theta}_{\bs{\xi}}$ does exist in the
form of a control polygon, $\CP_{k, \bs{\xi}, \bs{\theta}_{\bs{\xi}}},$ a strong convex hull for
$\bs{B}_{k, \bs{\xi}} \left(x \right) \bs{\theta}_{\bs{\xi}},$
\begin{equation}
  \label{eq:control_polygon}
  \CP_{k, \bs{\xi}, \bs{\theta}_{\bs{\xi}}} = \left\{ \left( \xi_{j}^{*},
  \theta_{j, \bs{\xi}} \right)
  \right\}_{j = 1}^{k + l}; \quad \xi_{j}^{*} = \frac{1}{k-1}
\sum_{i=1}^{k-1} \xi_{j + i}.
\end{equation}
$\CP_{k, \bs{\xi}, \bs{\theta}_{\bs{\xi}}}$ is a sequence of $k + l$ control
vertices.  We may interpret the control polygons as a piecewise linear function
approximating the spline function $\bs{B}_{k, \bs{\xi}} \left( x \right)
\bs{\theta}_{\bs{\xi}}.$ Changes in convexity and other subtle characteristics
of the spline function are exaggerated by the control polygon.  An example
basis, spline function, and control polygon are shown in
Figure~\ref{fig:basis_and_cp_plot}.  The control polygons are helpful for
illustrating the relationship between two splines with $\bs{\xi}_1 \subset
\bs{\xi}_2.$

\subsection{Relationship Between Splines of Different Dimensions}
Consider two knot sequences $\bs{\xi}$ and $\bs{\xi} \cup \bs{\xi}'.$
Then, for a given polynomial order $k,$ $\mathbb{S}_{k, \bs{\xi}} \subset
\mathbb{S}_{k, \bs{\xi}\cup \bs{\xi}'}$ \citep[pg 135]{deboor2001}.  Given this
relationship between spline spaces, and the convex sums generating spline
functions, \citet{boehm1980} presented a method for refining $\bs{\xi}$ without
affecting the spline function.  Specifically, for $\bs{\xi}$ and $\bs{\xi}'$
with $\card{\bs{\xi}'} = \sum_{x \in \bs{\xi}'} 1_{\left(
\min\left(\bs{\xi}\right), \max \left( \bs{\xi} \right) \right)} \left( x
\right),$ there exists an $\bs{\theta}_{\bs{\xi} \cup \bs{\xi}'}$ such that
$\bs{B}_{k, \bs{\xi}\cup\bs{\xi}'} \left( \bs{x} \right) \bs{\theta}_{\bs{\xi}
\cup \bs{\xi}'} = \bs{B}_{k, \bs{\xi}}
\left( \bs{x} \right) \bs{\theta}_{\bs{\xi}},$ for all $x \in \left[\min\left(
\bs{\xi} \right), \max \left( \bs{\xi} \right) \right].$

When refining $\bs{\xi}$ by the insertion of a singleton, $\xi',$ the
relationship between $\bs{\theta}_{\bs{\xi}}$ and $\bs{\theta}_{\bs{\xi} \cup
\xi'}$ is defined by the $\left(\card{\bs{\xi}} + 1\right) \times \card{\bs{\xi}}$ lower
bi-diagonal matrix 
\begin{equation}
  \label{eq:wmatrix}
  \bs{W}_{k, \bs{\xi}} \left( \xi' \right) =
  \begin{pmatrix}
    1                                          & 0                                          & \cdots                                                          & 0 \\
    1 - \omega_{1,k,\bs{\xi}}\left(\xi'\right) & \omega_{1,k,\bs{\xi}}\left(\xi'\right)     & \cdots                                                          & 0 \\
    0                                          & 1 - \omega_{2,k,\bs{\xi}}\left(\xi'\right) & \cdots                                                          & 0 \\
    \vdots                                     & \vdots                                     & \ddots                                                          & \vdots \\
    0                                          & 0                                          & 1 - \omega_{\card{\bs{\theta}} - 1,k,\bs{\xi}}\left(\xi'\right) & \omega_{\card{\bs{\theta}} - 1,k,\bs{\xi}}\left(\xi'\right) \\
    0                                          & 0                                          & 0                                                               & 1
  \end{pmatrix},
\end{equation}
where $\omega_{j,k,\bs{\xi}}(x)$ is as in \eqref{eq:omega} and
\begin{equation}
  \label{eq:boehm}
  \bs{\theta}_{\bs{\xi} \cup \xi'} = \bs{W}_{k, \bs{\xi}} \left( \xi' \right)
  \bs{\theta}_{\bs{\xi}}.
\end{equation}
Through recursion, \eqref{eq:boehm} can be used refine a knot sequences with the
insertion of multiple singletons without changing the values returned by the
spline function.

The $\bs{W}_{k, \bs{\xi}} \left( \xi' \right)$ matrix is full column rank. As
such, if given $\bs{\theta}_{\bs{\xi} \cup \xi},$ we can estimate
$\bs{\theta}_{\bs{\xi}}$ via the above relationship without having to refit a
full regression model.  This relationship is the foundation of our relative
influence weight metric.

\subsection{Relative Influence of Knots}

From \eqref{eq:boehm} we know that if $\xi_j \in \bs{\xi}$ has no influence on
$\bs{B}_{k, \bs{\xi}} \left( x \right) \bs{\theta}_{\bs{\xi}}$ then the regression coefficients $\bs{\theta}_{\bs{\xi} \backslash
\xi_j}$ are such that
\begin{equation}
  \bs{B}_{k, \bs{\xi}} \left( x \right) \bs{\theta}_{\bs{\xi}} = \bs{B}_{k, \bs{\xi}\backslash \xi_j} \left( x\right) \bs{\theta}_{\bs{\xi} \backslash \xi_j}
\quad \text{with} \quad
\bs{\theta}_{\bs{\xi}} = \bs{W}_{k, \bs{\xi}\backslash \xi_{j}}\left( \xi_{j}\right) \bs{\theta}_{\bs{\xi} \backslash \xi_{j}}.
\end{equation}
In practice we do not expect this equality to hold.  Instead, we expect to
observe, under the assumption that $\xi_{j}$ has no influence, 
\begin{equation}
  \bs{\theta}_{\bs{\xi}} = \bs{W}_{k, \bs{\xi}\backslash \xi_{j}}\left(\xi_{j}\right)
  \bs{\theta}_{\bs{\xi} \backslash \xi_{j}} + \bs{\epsilon}
\end{equation}
for some deviations $\bs{\epsilon}.$  As $\bs{\theta}_{\bs{\xi} \backslash
\xi_{j}}$ is unknown, we may estimate it via least squares, \ie,
\begin{equation}
  \bs{\theta}_{\bs{\xi} \backslash \xi_{j}} = 
  \left(
  \bs{W}^{T}_{k, \bs{\xi}\backslash \xi_{j}}\left(\xi_{j}\right)
  \bs{W}_{k, \bs{\xi}\backslash \xi_{j}}\left(\xi_{j}\right)
  \right)^{-1}
  \bs{W}^{T}_{k, \bs{\xi}\backslash \xi_{j}}\left(\xi_{j}\right) 
  \bs{\theta}_{\bs{\xi}}.
\end{equation}

\begin{figure}
  \centering
\begin{Schunk}

\includegraphics[width=\maxwidth]{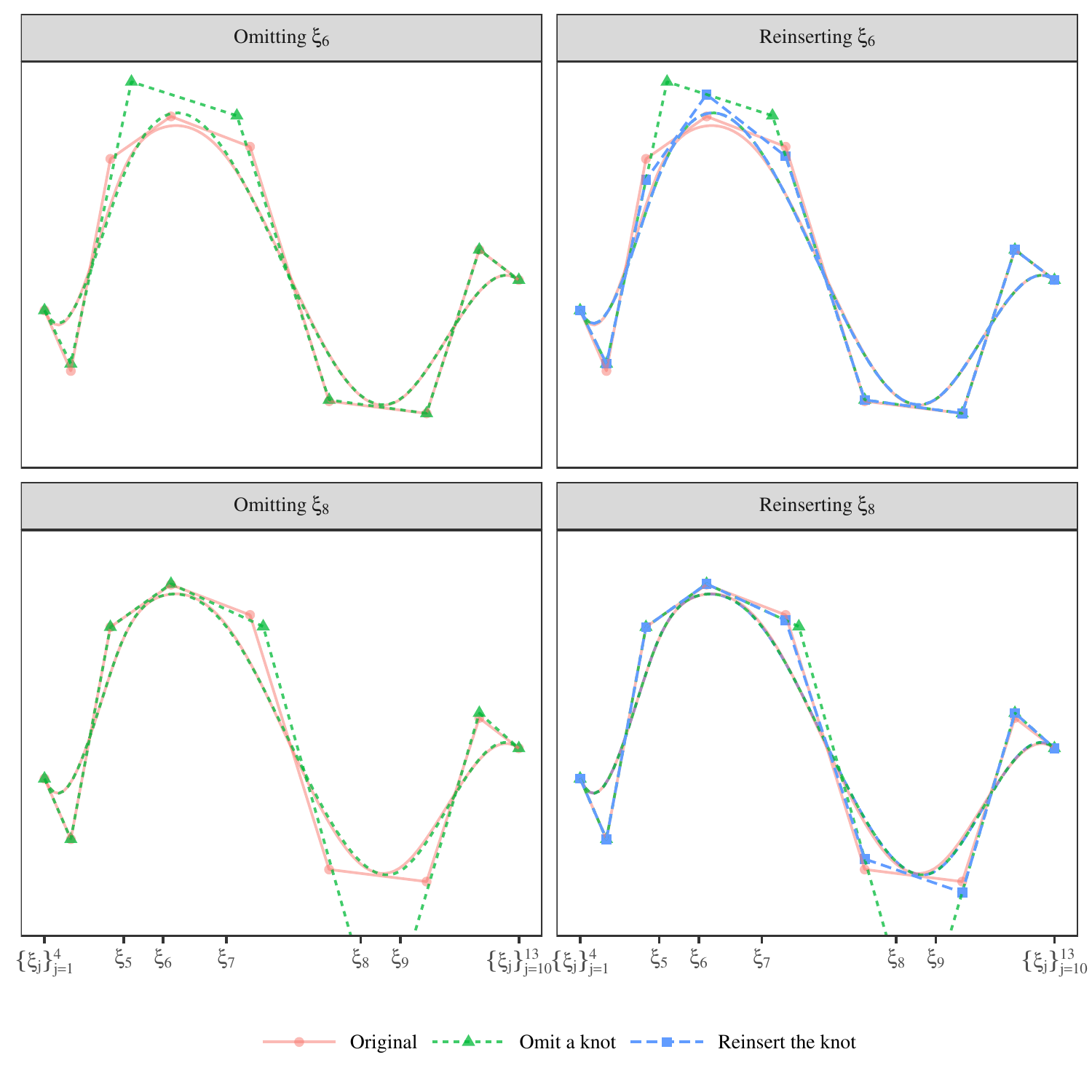} \end{Schunk}
  \caption[Examples of knot influence on a spline function.]{%
    Two examples of omitting and reinserting a knot to determine the influence
    of the knot on the spline function.  The original control polygon is as in
    Figure~\ref{fig:cp_plot}.  The top row of plots here illustrate the
    influence of $\xi_6$ and the bottom row illustrate the influence of $\xi_8.$
    In the left column we present the original control polygon $\CP_{k,
    \bs{\xi}, \bs{\theta}_{\bs{|xi}}}$ and the
    control polygon based on a coarsened knot sequence $\CP_{k, \bs{\xi}
    \backslash \xi_j, \bs{\theta}_{\bs{\xi} \backslash \xi_j}}.$  The right column shows
    the original control polygon, the coarsened control polygon, and the control
    polygon after reinsertion of $\xi_j,$ $\CP_{k, \left(\bs{\xi} \backslash \xi_j\right) \cup \xi_j, \bs{\theta}_{\left(\bs{\xi} \backslash \xi_j\right) \cup \xi_j}}.$  The influence
    weights of $\xi_6$ and $\xi_8$ are 0.539 and
    0.278 respectively.  
  }
  \label{fig:influence_weight_plots}
\end{figure}

As illustrated in Figure~\ref{fig:influence_weight_plots} the spline initial
spline $\bs{B}_{k, \bs{\xi}}\left(x\right) \bs{\theta}_{\bs{\xi}}$ and
$\bs{B}_{k, \bs{\xi}\backslash \xi_j}\left( x \right) \bs{\theta}_{\bs{\xi}
\backslash \xi_j}$ will differ.  The control polygons help to elucidate the
differences between these two splines.  A measure of the difference between the
two splines or between the two control polygons is needed to assess the
influence of $\xi_{j}.$  If we `reinsert' $\xi_{j}$ into $\bs{\xi} \backslash
\xi_{j}$ then we can get a vector of ordinates $\bs{\theta}_{\left(\bs{\xi}
\backslash \xi_{j} \right) \cup \xi_{j}}$ to approximate $\bs{\theta}_{\xi}.$
Again using \eqref{eq:boehm} we find the values of $\bs{\theta}_{\left(\bs{\xi}
\backslash \xi_{j} \right) \cup \xi_j},$ \begin{equation} \begin{split}
  \bs{\theta}_{\left( \bs{\xi} \backslash \xi_{j} \right) \cup \xi_{j}} &=
  \bs{W}_{k, \bs{\xi}\backslash \xi_{j}}\left(\xi_{j}\right)
  \bs{\theta}_{\bs{\xi} \backslash \xi_{j}} \\ &= \bs{W}_{k, \bs{\xi}\backslash
\xi_{j}}\left(\xi_{j}\right) \left( \bs{W}^{T}_{k, \bs{\xi}\backslash
  \xi_{j}}\left(\xi_{j}\right) \bs{W}_{k, \bs{\xi}\backslash
  \xi_{j}}\left(\xi_{j}\right) \right)^{-1} \bs{W}^{T}_{k, \bs{\xi}\backslash
\xi_{j}}\left(\xi_{j}\right) \bs{\theta}_{\bs{\xi}}.  \end{split} \end{equation}

By this construction, 
$\bs{B}_{k, \bs{\xi}\backslash \xi_j]}\left(x\right)\bs{\theta}_{\bs{\xi} \backslash \xi_j} = 
\bs{B}_{k, \left(\bs{\xi}\backslash \xi_j\right) \cup \xi_j]}\left(x\right)\bs{\theta}_{\left(\bs{\xi} \backslash \xi_j\right) \cup \xi_j}.$
The control polygons
$\CP_{k, \bs{\xi}, \bs{\theta}_{\bs{\xi}}}$ and $\CP_{k, \left(\bs{\xi} \backslash \xi_{j} \right) \cup
\xi_{j}, \bs{\theta}_{\left(\bs{\xi} \backslash \xi_j \right) \cup \xi_j}}$ have the same abscissae but differ in their ordinates.
The squared length of the residual vector, \ie, the squared Euclidean distance between the ordinates
$\bs{\theta}_{\bs{\xi}}$ and $\bs{\theta}_{\left( \bs{\xi} \backslash \xi_j
\right) \cup \xi_j}$ is the influence weight for $\xi_{j}.$  That is, 
the influence weight of $\xi_j \in \bs{\xi}$ for $\CP$ is
\begin{equation}
  \label{eq:influence_weight}
  \begin{split}
    w_j
    &= \left\lVert \bs{\theta}_{\bs{\xi}} - \bs{\theta}_{\left( \bs{\xi} \backslash \xi_j
    \right) \cup \xi_j} \right\rVert_{2} \\
    &=
    \left\lVert 
    \left(\bs{I} - 
    \bs{W}_{k, \bs{\xi}\backslash \xi_{j}}\left(\xi_{j}\right) 
    \left(
    \bs{W}^{T}_{k, \bs{\xi}\backslash \xi_{j}}\left(\xi_{j}\right)
    \bs{W}_{k, \bs{\xi}\backslash \xi_{j}}\left(\xi_{j}\right)
    \right)^{-1}
    \bs{W}^{T}_{k, \bs{\xi}\backslash \xi_{j}}\left(\xi_{j}\right) 
    \right)
    \bs{\theta}_{\bs{\xi}}
    \right\rVert_{2}.
  \end{split}
\end{equation}

The influence weights $w_j$ are non-negative and provide a relative measure of
the influence of knots with in a sequence.  See
Figure~\ref{fig:influence_weight_plots} to compare the relative influence of
$\xi_6$ and $\xi_8$ on the spline.
Knot  $\xi_8$ has a smaller influence
weight, 
$w_8 = 0.278,$  
than $\xi_6,$
$w_6 = 0.539.$ 
As such, if we were required to reduce the
cardinality of $\bs{\xi}$ by one, then omission of $\xi_8$ would be preferable
to the omission of $\xi_6$ because $\xi_8$ has a lower influence on the spline
function than $\xi_6.$

\section{Control Polygon Reduction} \label{sec:cpr_algorithm}

CPR is based on the assumption that $f$ in~\eqref{eq:cpr_general_model} can be
sufficiently modeled by $\bs{B}_{k, \bs{\xi}} \left(x\right)
\bs{\theta}_{\bs{\xi}} \in \mathbb{S}_{k, \bs{\xi}}.$ If we start searching for $\bs{B}_{k, \bs{\xi}}
\left(x\right) \bs{\theta}_{\bs{\xi}}$ in a larger space $\mathbb{S}_{k, \bs{\xi} \cup \bs{\xi}'}$
then there must be at least one internal
knot within $\bs{\xi} \cup \bs{\xi}'$ which is unnecessary to sufficiently model $f.$

By starting with a large knot sequence, $\card{\bs{\xi}} - 2k = L$ internal
knots, is not without
precedence.  Large quantities of internal knots have been used, and strongly encouraged
by \citet{eilers1996, eilers2010} and \citet{binder2008b}.  While the B-spline
models are sensitive to knot location, the difference between a location of
$\xi_j$ and $\xi_j + \varepsilon,$ for some small $\left|\varepsilon \right| >
0,$ is negligible.

CPR is a backward-step model selection algorithm---we start with a large
degree of freedom model and reduce the degrees of freedom incrementally.  Final
selection of the model is not done by an automated stopping criterion.  Instead,
sequentially larger models, starting with the zero internal knot model, are
assessed by fit statistics or visual inspection until the analyst is satisfied
that an additional knot provides only negligibly better fit.

As a model selection method, CPR is preferable to likelihood based forward-step,
backward-step, and grid search methods.   First, a forward-step method requires
placing one knot somewhere on the support, generally on the median, then placing
two knots, generally on the tertiles, and so forth.  The model space is limited
to only one model of each dimension.  Further, sequential models are are not
within the same spline space.  This can make finding an acceptable model difficult
as higher dimensional models may be selected so that a needed low dimensional
knot sequence is used within the larger knot sequence.  CPR allows for the location of
interior knots to be selected, that is, a one knot model may have the knot at
any location, depending on initial knot sequence construction.  The difference
computational time required to run CPR versus a forward-step process is
negligible.

Using a grid search of possible knot locations is computationally impractical.
For $L$ possible initial knot locations there are $\sum_{l=0}^{L} \binom{L}{l} =
2^L$ possible models to fit.  With $L = 20$ possible knot locations there would
be $1,048,576$ regression models.
Likely more models than data observations, and simply too computationally
expensive to be done.

Using a backward-step likelihood based selection method requires far fewer
regression model fits.  For $L$ possible knot locations there are $L(L+1)/2 + 1$
regression models to fit, that is, 
$211$ models for twenty possible
locations,
$1,276$ models for fifty possible
locations, and $5,051$ models for
one-hundred possible locations.  The computation time can still be considerable, but at
least the backward-step approach is reasonable, compared to the grid search.

CPR is similar to the backward-step likelihood based approaches, but requires
only $L + 1$ regression models to be fitted for a set of $L$ initial knot
locations.  The computation time saved by requiring only $L + 1$ versus
$L(L+1)/2 + 1$ can be considerable.

\paragraph{The CPR Algorithm}
\begin{enumerate}
  \item Start with a knot sequence with a sufficiently large number of interior
    knots, say $L = 50,$ and set $l = L$ to index models.
  \item Use an appropriate regression modeling approach to fit a regression
    model for~\eqref{eq:cpr_general_model}. 
  \item \label{step:build_cp} Construct the control polygon for the current
    $\bs{\xi}^{(l)}$ and $\bs{\theta}^{(l)}$ estimate.
  \item\label{step:weights} Use $\CP_{k, \bs{\xi}^{(l)}, \bs{\theta}^{(l)}}$ and
    \eqref{eq:influence_weight} to find the influence weight for all internal
    knots.
  \item\label{step:coarsen} Coarsen the knot sequence by removing the knot with
    the smallest influence weight.
  \item\label{step:refit} Refit the regression model using the coarsened knot
    sequence and index $l = l - 1.$
  \item Repeat steps \ref{step:build_cp} through \ref{step:refit} until all
    internal knots have been removed, {\itshape i.e.}, if $l \geq 0$ goto
      \ref{step:build_cp}, else goto \ref{step:selection}.
  \item\label{step:selection} Select the preferable model by visual inspection
    of diagnostic graphics.
\end{enumerate}

There are two diagnostic plots we suggest using for step \ref{step:selection}.
1) Consider sequential control polygons.  Starting with $l = 0$ internal knots,
compare the control polygon for $l$ and $l + 1$ internal knots.  When the
control polygon with $l+1$ internal knots appears to be nested within the
control polygon based on $l$ internal knots, the preferable model is the one
with $l$ internal knots.  2) Plot the root mean squared error (RMSE) as a
function of the number of internal knots.  When there is no longer a meaningful
decrease in the RMSE between models with $l$ and $l + 1$ internal knots, select
the model with $l$ internal knots.  These diagnostic plots will be shown in
examples below.

\citet{lyche1988b} presented a similar grouped knot removal process when
estimating a function with no noise via B-splines.  CPR differs from Lyche and
M{\o}rken in four key ways: 1) CPR is applicable to noisy data with an unknown
target function whereas Lyche and M{\o}rken had a known target function.  2) The
regression models selected by CPR will approximate $f(x) \approx \bs{B}_{k,
\bs{\xi}} \left(x \right) \bs{\theta}$ whereas Lyche and M{\o}rken made
estimates of the form $f \left( \xi_{i}^{*} \right) \approx \theta_i.$ 3) CPR
coarsens the knot sequence by removal of one element and then refits the
regression model.  Lyche and M{\o}rken coarsen the knot sequence via removal of
multiple knots before refitting the regression model.  4) Given that the target
function is unknown, CPR has no automated stopping criteria.

\section{The cpr Package} \label{sec:cpr_package} 
The objective of the \pkg{cpr} is to provide a simple, intuitive, and clean API
for the CPR algorithm.  A four-line script is the foundation of the expected use
case.
\begin{Schunk}
\begin{Sinput}
R> initial_cp    <- cp(y ~ bsplines(x), data = your_data_frame)
R> cpr_run       <- cpr(initial_cp)
R> plot(cpr_run)
R> preferable_cp <- cpr_run[[4]]
\end{Sinput}
\end{Schunk}
In the first line the initial control polygon is constructed via a common
regression statement.  Passing the initial control polygon object to the
\code{cpr} function applies the CPR algorithm and returns a list of control
polygons in the \code{cpr_run} object.  The \code{plot} call generates the
diagnostic plots used for selecting a preferable model.  Lastly, the preferable
control polygon is retrieved.

In the following subsections we will present details on the relevant calls noted
above.  We start with \code{bsplines} for generating B-spline bases and follow
with \code{cp} for construction of control polygons.  We will then illustrate
the use of \code{cpr} for selecting good fitting parsimonious B-spline
regression models.

\pkg{cpr} relies on \pkg{Rcpp}~\citep{eddelbuettel2011rcpp,
eddelbuettel2013seamless} and
\pkg{RcppArmadillo}~\citep{eddelbuettel2014rcpparmadillo} to gain computational
efficiency for several functions and matrix arithmetic.  Heavy reliance on
\pkg{dplyr}~\citep{wickham2016dplyr} and \pkg{tidyr}~\citep{wickham2016tidyr}
for data manipulation.  Two-dimensional graphics are generated via
\pkg{ggplot2}~\citep{wickham2009ggplot2}. Three-dimensional graphics are
generated using either \pkg{rgl}~\citep{adler2016rgl} for dynamic iterative
graphics, or \pkg{plot3D}~\citep{r-plot3D} of static graphics.

\subsection{B-Splines} \label{sec:cpr_bs}
The standard installation of \proglang{R} includes the
\pkg{splines}~\citep{R-base} package and the \code{splines::bs} function for generating
the basis matrix of B-splines, {\it i.e.} the matrix shown in
\eqref{eq:basis_matrix}.  The \pkg{cpr} package provides an alternative
function, \code{cpr::bsplines} for generating B-spline basis matrices with the class
\code{cpr\_bs}.  The differences in the functional arguments and the attributes
of the return objects between \code{splines::bs} and \code{cpr::bsplines} are listed in
Table~\ref{tab:bs_vs_bsplines}. 
\begin{table}
  \caption{Comparison of the arguments, with default values, for \code{splines::bs} and
    \code{cpr::bsplines}.  The attributes for the resulting \code{splines::bs} and
  \code{cpr\_bs} objects are also reported.}
  \label{tab:bs_vs_bsplines}
  \centering
\begin{tabular}{lll} \hline
           & \code{splines::bs}               & \code{cpr::bsplines}     \\ \hline
 Arguments &                                  &                          \\
           & \code{x}                         & \code{x}                 \\
           & \code{df}                        & \code{df}                \\
           & \code{knots}                     & \code{iknots}            \\
           & \code{degree = 3}                & \code{order = 4L}        \\
           & \code{Boundary.knots = range(x)} & \code{bknots = range(x)} \\
           & \code{intercept = FALSE}         & --                       \\ \hline
Attributes &                                  &                          \\
           & \code{dim}                       & \code{dim}               \\
           & \code{degree}                    & \code{order}             \\
           & \code{knots}                     & \code{iknots}            \\
           & \code{Boundary.knots}            & \code{bknots}            \\
           & \code{intercept}                 & --                       \\
           & --                               & \code{xi}                \\
           & --                               & \code{xi\_star}          \\
           & \code{class}                     & \code{class}             \\
           & --                               & \code{call}              \\
           & --                               & \code{environment}       \\ \hline
\end{tabular}
\end{table}

A major difference between the two functions is related to the \code{intercept}
argument of \code{splines::bs}.  By default, \code{splines::bs} will omit the
first column of the basis whereas \code{cpr::bsplines} will return the whole
basis.  The omission of the first column of the basis generated by
\code{splines::bs} allows for additive \code{splines::bs} calls to be used on the
right-hand-side of a regression formula and generate a full rank design matrix.
If additive \code{cpr::bsplines} calls, or additive \code{splines::bs} with \code{intercept =
TRUE}, are on the right-hand-side of the regression equation the resulting
design matrix will be rank deficient.  This is a result of the B-splines being a
partition of unity.  As the CPR algorithm is based on having the whole basis,
the \code{cpr::bsplines} function is provided to make it easy to work with the whole
basis without having to remember to use non-default settings in \code{splines::bs}.
The default call \code{splines::bs(x)} is replicated
by \code{cpr::bsplines(x)[, -1]} and the default call \code{cpr::bsplines(x)} is
replicated \code{splines::bs(x, intercept = TRUE)}.

Specifying the polynomial order and knot sequence between the two functions
differ between \code{splines::bs} and \code{cpr::bsplines}.
\code{splines::bs} uses the polynomial \code{degree} whereas \code{cpr::bsplines} uses the
polynomial \code{order} (order = degree + 1) to define the splines.  The default
for both \code{splines::bs} and \code{cpr::bsplines} is to generate cubic B-splines.

For both \code{splines::bs} and
\code{cpr::bsplines} only the degrees of freedom or the internal knots need to be
specified.  If the end user specifies both, the specified knots take precedence.
If only \code{df} is specified then \code{df - order} internal knots will be
generated.
\code{splines::bs} and \code{cpr::bsplines}.
For a numeric vector \code{x}, 
\code{splines::bs} will generate a sequence of
internal knots via a call equivalent to 
\begin{Schunk}
\begin{Sinput}
R> knots <- df + order + (1L - intercept)
R> stats::quantile(x,
+    probs = seq(0, 1, length = length(knots) + 2L)[-c(1, length(knots) + 2L)])
\end{Sinput}
\end{Schunk}
whereas \code{cpr::bsplines} will generate a sequence  equivalent to
\begin{Schunk}
\begin{Sinput}
R> stats::quantile(unique(x)[-c(1, length(unique(x)))],
+                  probs = seq(1, df - order, by = 1) / (df - order + 1))
\end{Sinput}
\end{Schunk}
The function \code{cpr::trimmed_quantile} is provided to generate such
sequences.

The return object from both \code{splines::bs} and \code{cpr::bsplines} is a matrix.  The
attributes returned include the argument values used to construct the basis.
The major difference in the attributes between the two objects is that
\code{cpr::bsplines} returns the full knot sequence, $\bs{\xi},$ in the \code{xi}
element and the Greville sites, $\bs{\xi}^{*}$ in the \code{xi\_star} element.
These attributes are used in the construction of control polygons.
Lastly, the classes for the two objects differ: \code{splines::bs} returns a three
classes, \code{c("bs", "basis", "matrix")} and \code{cpr::bsplines} returns two
classes, \code{c("cpr\_bs", "matrix")}.  An example construction and structure
are below.
\begin{Schunk}
\begin{Sinput}
R> bmat <- bsplines(x      = seq(0, 6, length = 500),
+                   iknots = c(1.0, 1.5, 2.3, 4.0, 4.5))
R> bmat
\end{Sinput}
\begin{Soutput}
Matrix dims: [500 x 9]

      [,1]   [,2]     [,3]     [,4] [,5] [,6] [,7] [,8] [,9]
[1,] 1.000 0.0000 0.000000 0.00e+00    0    0    0    0    0
[2,] 0.964 0.0354 0.000287 5.04e-07    0    0    0    0    0
[3,] 0.930 0.0693 0.001137 4.03e-06    0    0    0    0    0
[4,] 0.896 0.1018 0.002537 1.36e-05    0    0    0    0    0
[5,] 0.863 0.1330 0.004471 3.22e-05    0    0    0    0    0
[6,] 0.830 0.1627 0.006924 6.30e-05    0    0    0    0    0
\end{Soutput}
\end{Schunk}

There is no default method for plotting \code{splines::bs} objects.  If the numeric
vector \code{x} is sorted, then a minimally useful basis plot can be generated
via \code{graphics::matplot}.  The \pkg{cpr} package provides a plotting method
for the \code{cpr\_bs} objects. The plotting method returns a \code{c("gg",
"ggplot")} object and can be modified by adding additional layers as would be
done for any other \code{ggplot} object.  For example, the basis plot in
Figure~\ref{fig:basis_plot} was generated by
\begin{Schunk}
\begin{Sinput}
R> plot(bmat, show_xi = TRUE, show_x = TRUE,  color = TRUE, digits = 1) +
+  theme(text = element_text(family = "Times", size = 10))
\end{Sinput}
\end{Schunk}

\subsection{Control Polygons} \label{sec:cpr_cp}
\code{cpr\_cp} objects are constructed by the S3 generic function
\begin{Schunk}
\begin{Sinput}
R> methods("cp")
\end{Sinput}
\begin{Soutput}
[1] cp.cpr_bs*  cp.formula*
see '?methods' for accessing help and source code
\end{Soutput}
\end{Schunk}

The \code{cpr:::cp.cpr\_bs} method take a \code{cpr\_bs} object and a vector of
ordinates to construct a control polygon as defined
in~\eqref{eq:control_polygon}.  For example, the control polygon showing in
Figure~\ref{fig:cp_plot} is generated by the \code{bmat} above and the following
ordinates.
\begin{Schunk}
\begin{Sinput}
R> bmat <- bsplines(x      = seq(0, 6, length = 500),
+                   iknots = c(1.0, 1.5, 2.3, 4.0, 4.5))
R> theta <- c(1, 0, 3.5, 4.2, 3.7, -0.5, -0.7, 2, 1.5)
R> eg_cp <- cp(bmat, theta)
R> str(eg_cp)
\end{Sinput}
\begin{Soutput}
List of 10
$ cp :Classes 'tbl_df', 'tbl' and 'data.frame': 9 obs. of 2 variables:
..$ xi_star: num [1:9] 0 0.333 0.833 1.6 2.6 ...
..$ theta : num [1:9] 1 0 3.5 4.2 3.7 -0.5 -0.7 2 1.5
$ xi : num [1:13] 0 0 0 0 1 1.5 2.3 4 4.5 6 ...
$ iknots : num [1:5] 1 1.5 2.3 4 4.5
$ bknots : num [1:2] 0 6
$ order : num 4
$ call : language cp.cpr_bs(x = bmat, theta = theta)
$ keep_fit: logi NA
$ fit : logi NA
$ loglik : logi NA
$ rmse : logi NA
- attr(*, "class")= chr [1:2] "cpr_cp" "list"
\end{Soutput}
\end{Schunk}
The resulting \code{cpr\_cp} object seems trivial.  A \code{data.frame} with the
Greville sites and the given ordinates, along with attributes of the B-spline
basis.  There are several elements related to regression model fits which are
\code{NA} when a \code{cpr.cpr_bs} is used to generate the control polygon.
The regression related elements of a \code{cpr\_cp} object are populated when
the control polygon is generated using the \code{cpr:::cp.formula} method.

The \code{cp.formula} method is one of the most useful functions provided in
\pkg{cpr}.  This method uses regression methods to determine the control
vertices and is called many times when running the CPR algorithm.
\begin{Schunk}
\begin{Sinput}
R> str(cpr:::cp.formula)
\end{Sinput}
\begin{Soutput}
function (formula, data, method = stats::lm, ..., keep_fit = FALSE, check_rank =
   TRUE)
\end{Soutput}
\end{Schunk}
The arguments for this function are
\begin{itemize}
  \item \code{formula} a regression formula, sufficient for the regression
    \code{method} which contains one call to \code{bsplines} on the
    right-hand-side.
  \item \code{data} a required \code{data.frame} containing the variables noted
    in the \code{formula}.
  \item \code{method} the regression method.  By default this is the \code{lm}
    call from the base \code{stats} package.  This regression method is used to
    estimate the ordinates for the control polygon.
  \item \code{\ldots} additional arguments passed to \code{method}.
  \item \code{keep\_fit} If \code{TRUE} the object returned from \code{method}
    is stored in the resulting \code{cpr\_cp} object.  When \code{FALSE} only
    some summary statistics are retained.
  \item \code{check\_rank} checks that the design matrix for the regression
    model is sufficient.  This is done via the \code{cpr::matrix\_rank}
    function.
\end{itemize}

A simple example, fitting the sine function with a cubic B-spline with ten
internal knots, is below.
\begin{Schunk}
\begin{Sinput}
R> dat <- tibble::data_frame(x = seq(-pi, pi, length = 200),
+                            y = sin(x))
R> eg_cp2 <- cp(y ~ bsplines(x = x, df = 14), data = dat)
\end{Sinput}
\end{Schunk}

A quick note about how \code{cp.formula} generates design matrices.  By
construction, full B-spline basis
matrix, a partition of unity, is returned by \code{cpr::bsplines}. The standard intercept in a
regression model, implicit in \proglang{R} regression formulae, would be
co-linear with the basis and a rank deficient design matrix would be generated.
To elevate this issue, \code{cp.formula} automatically prepends the
right-hand-side of the \code{formula} with \code{0 +} to omit the intercept.
This is also consistent with \eqref{eq:cpr_general_model}, a varying means
model.  It is also worth noting that additional additive terms can be added to
the right-hand-side of the \code{formula}.  The code{cp.formula} call will
correctly build design matrices for model formulae when additive continuous
variables and/or categorical variables are specified.

\subsection{Relative Influence of the Knots}

Our metric for assessing the relative influence of a knot was derived via the
geometry of control polygons. Thus, \pkg{cpr} provides the
\code{cpr::influence_of} function to calculate and provide
the relative influence weight for each internal knot of a given \code{cpr_cp} object.

\begin{Schunk}
\begin{Sinput}
R> influence_of(eg_cp)
\end{Sinput}
\begin{Soutput}
# A tibble: 5 × 4
  index iknots     w  rank
  <int>  <dbl> <dbl> <dbl>
1     5    1.0 1.283     5
2     6    1.5 0.539     2
3     7    2.3 0.559     3
4     8    4.0 0.278     1
5     9    4.5 0.648     4
\end{Soutput}
\end{Schunk}

The default behavior of \code{influence_of} is to return the influence weight
of all internal knots.  The end user may request only a subset of knots be
evaluated by passing the \code{indices}\footnote{\code{indices} based on
the full knot sequence, not just the interior knots.  That is, for a fourth
order spline, the first interior knot is index 5 as knots 1:4 are the left
boundary knots.} of the knot of interest as the second argument to
\code{influence_of}.

The output printed to the console gives the index of each knot, the value of the
knot, weight, and rank with rank == 1 given to the least influential.  The
structure of the returned \code{cpr_influence_of} object is more complex than
the output belies.

\begin{Schunk}
\begin{Sinput}
R> str(influence_of(eg_cp), max.level = 1)
\end{Sinput}
\begin{Soutput}
List of 5
$ weight :Classes 'tbl_df', 'tbl' and 'data.frame': 5 obs. of 4 variables:
$ orig_cp :List of 10
..- attr(*, "class")= chr [1:2] "cpr_cp" "list"
$ indices : int [1:5] 5 6 7 8 9
$ coarsened_cp :List of 5
$ reinserted_cp:List of 5
- attr(*, "class")= chr "cpr_influence_of"
\end{Soutput}
\end{Schunk}

The \code{weight} \code{data.frame} is what is shown in the console.  The
original control polygon and a lists of each coarsened and reinserted control
polygon are also returned.  These lists of control polygons are useful for
plotting the results.  A plot similar to the right column of
Figure~\ref{fig:influence_weight_plots} can be generated via:

\begin{Schunk}
\begin{Sinput}
R> plot(influence_of(eg_cp, c(6, 8)))
\end{Sinput}
\end{Schunk}

\subsection{Model Selection via Control Polygon Reduction}
Here we demonstrate model selection via the CPR algorithm.  
In the
following example we will use the \code{spdg} data set provided in the \pkg{cpr}
package to model the progesterone hormone profile expressed during a menstrual
cycle.  The \code{spdg} data set is a simulated data set based on summary
statistics of a subset
of the Study
of Women's Health Across the Nation (SWAN) Daily Hormone Study
(DHS)~\citep{santoro2003}.  SWAN is a ``multi-site longitudinal, epidemiologic
study designed to examine the health of women during their middle years.''  The
DHS was a specific sub-study in which subject provided first evacuation urine
samples every day for a full menstrual cycle.  Pregnanediol glucuronide (PDG),
the urine metabolite of progesterone, was one of four reproductive hormones
measured from the urine samples. The summary statistics and script used to
generate the simulated data set can be found in the github repository for the
\pkg{cpr} package, \url{https://github.com/dewittpe/cpr/tree/master/data-raw}.

\begin{Schunk}
\begin{Sinput}
R> str(spdg)
\end{Sinput}
\begin{Soutput}
Classes 'tbl_df', 'tbl' and 'data.frame':	24730 obs. of  9 variables:
$ id : int 1 1 1 1 1 1 1 1 1 1 ...
$ age : num 49.3 49.3 49.3 49.3 49.3 ...
$ ttm : num -5.19 -5.19 -5.19 -5.19 -5.19 ...
$ ethnicity : Factor w/ 5 levels "Caucasian","Black",..: 4 4 4 4 4 4 4 4 4 4 ...
$ bmi : num 36.2 36.2 36.2 36.2 36.2 ...
$ day_from_dlt: num -8 -7 -6 -5 -4 -3 -2 -1 0 1 ...
$ day_of_cycle: int 1 2 3 4 5 6 7 8 9 10 ...
$ day : num -1 -0.875 -0.75 -0.625 -0.5 ...
$ pdg : num 0.2401 0.0668 0.1088 0.0733 0.0979 ...
\end{Soutput}
\begin{Sinput}
R> dplyr::n_distinct(spdg$id)
\end{Sinput}
\begin{Soutput}
[1] 864
\end{Soutput}
\end{Schunk}

\begin{figure}
  \centering
\begin{Schunk}

\includegraphics[width=\maxwidth]{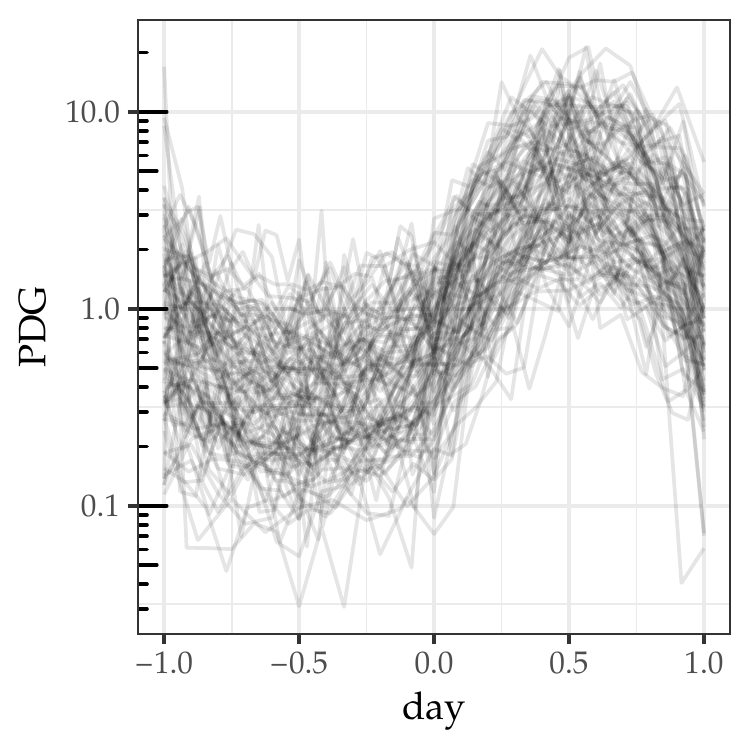} \end{Schunk}
\caption{Simulated PDG data for 100 subjects in the \code{spdg} data set.}
\label{fig:spdg_plot}
\end{figure}

The \code{spdg} data set contains \code{pdg} values for one full cycle from
864 subjects.  Subject level variables are
\code{age} in years, time-to-menopause, \code{ttm}, for years before reaching
menopause, \code{ethnicity}, and body-mass-index, \code{bmi}.
\code{day_of_cycle} are positive integers and \code{day_from_dlt} give the day
away from the day of luteal transition (DLT).  The DLT is the day between the
follicular phase and luteal phase of the cycle.  \code{day_from_dlt == 0} is the
DLT, negative values are for the follicular phase, and positive values for the
luteal phase.  \code{day} is a mapping of \code{day_from_dlt} to $[-1, 1]$ with
\code{day == 0} being the DLT.  The follicular and luteal phases are mapped to
$[-1, 0)$ and $(0, 1]$ respectively via a linear mapping.  Lastly, the simulated
PDG values are given in the \code{pdg} element.  Figure~\ref{fig:spdg_plot}
shows the $\log_{10}\left( \text{\code{pdg}} \right)$ values by \code{day} for
the \code{spdg} data set.

\begin{figure}
  \centering
  \begin{subfigure}[t]{0.48\linewidth}
\begin{Schunk}

\includegraphics[width=\maxwidth]{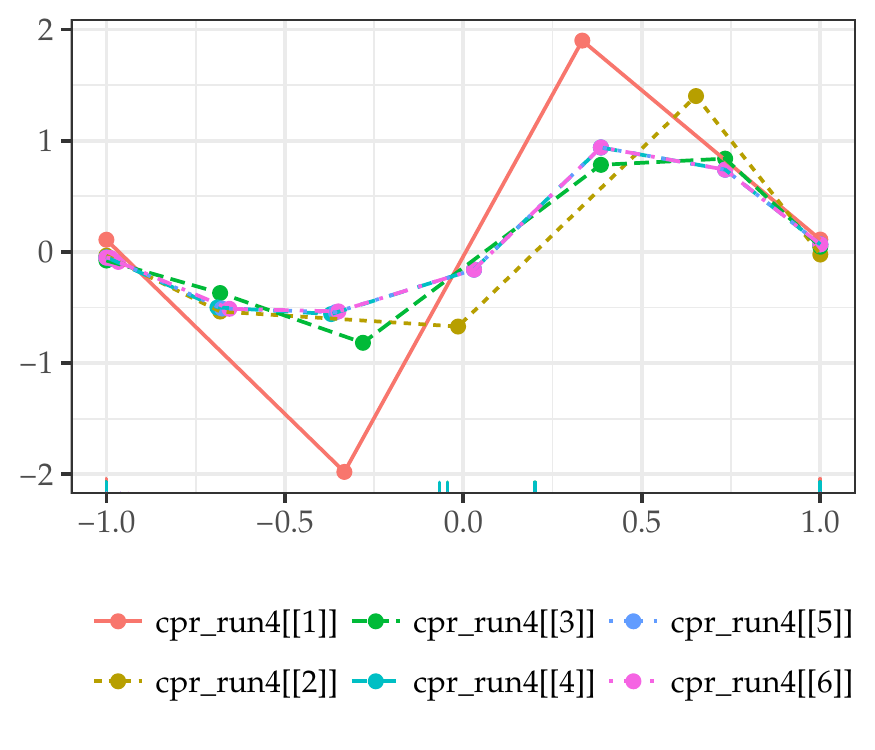} \end{Schunk}
    \caption{}
    \label{fig:cpr_run4_plot1}
  \end{subfigure}
  ~
  \begin{subfigure}[t]{0.48\linewidth}
\begin{Schunk}

\includegraphics[width=\maxwidth]{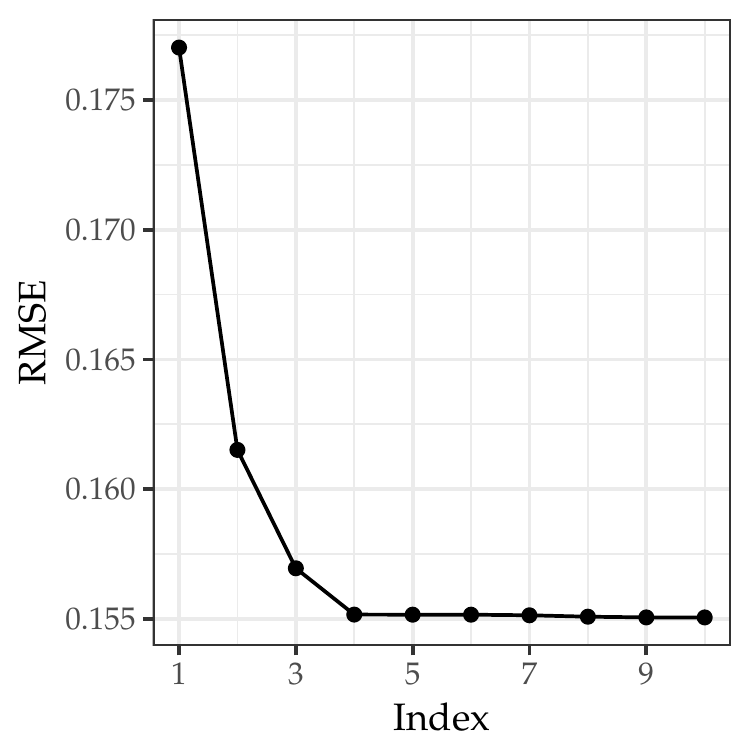} \end{Schunk}
    \caption{}
    \label{fig:cpr_run4_plot2}
  \end{subfigure}
  \caption{CPR diagnostic plots.  (a) show the sequential control polygons.
    Noticeable differences between the control polygons exist from index 1, 2,
    3, and 4.  The differences between \code{cpr\_run4[[4]]},
    \code{cpr\_run4[[5]]}, and \code{cpr\_run4[[6]]} are almost indistinguishable.
    Ergo, \code{cpr\_run4[[4]]} is the preferable model. (b) show the root mean
    squared error (RMSE) by model index.  A meaningful decrease in RMSE occurs
    with additional degrees of freedom until we look at the change between model
    index 4 and 5.  Model index 4 is the preferable model.}
  \label{fig:cpr_run4_plots}
\end{figure}

We will start our search for a parsimonious B-spline regression model with a
high quality of fit with fourth order B-splines and fifty internal knots. We will
fit a linear mixed model via \code{lmer} from the \pkg{lme4} package~\citep{r-lme4}.
\begin{Schunk}
\begin{Sinput}
R> initial_cp4 <- cp(log10(pdg) ~ bsplines(day, df = 54) + (1|id),
+                    data = spdg, method = lmer) 
\end{Sinput}
\end{Schunk}
Applying the CPR algorithm requires one call to \code{cpr}.
\begin{Schunk}
\begin{Sinput}
R> cpr_run4 <- cpr(initial_cp4)
R> cpr_run4
\end{Sinput}
\begin{Soutput}
A list of control polygons
List of 51
- attr(*, "class")= chr [1:2] "cpr_cpr" "list"
\end{Soutput}
\end{Schunk}
The \code{cpr_run4} object is a list of \code{cpr_cp} objects.  Index $i$ is based
on a control polygon with $i - 1$ internal knots.  There are two plots that can
be used for selection of a preferable model.  The \code{plot} method returns a
\code{ggplot} object and can be modified accordingly.
\begin{Schunk}
\begin{Sinput}
R> plot(cpr_run4, color = TRUE) + theme(legend.position = "bottom") # Figure \ref{fig:cpr_run4_plot1}
R> plot(cpr_run4, type = "rmse", to = 10) + 
+  ylab("RMSE") +
+  scale_x_continuous(breaks = seq(1, 10, by = 2))   # Figure \ref{fig:cpr_run4_plot2}
\end{Sinput}
\end{Schunk}

Recall that if $\xi_j$ has no influence on a spline, then the vertices of
$\CP_{k, \bs{\xi}, \bs{\theta}_{\bs{\xi}}}$ will be on the edges of $\CP_{k,
\bs{\xi} \backslash \xi_j, \bs{\theta}_{\bs{\xi} \backslash \xi_j}}.$  In
Figure~\ref{fig:cpr_run4_plot1} we conclude that \code{cpr_run4[[4]]} is the
preferable model as the control polygons in index 4, 5, and 6 are
indistinguishable.  The same conclusion, based on decrease in RMSE by model
index, is seen in Figure~\ref{fig:cpr_run4_plot2}.

After selecting the preferable model you can use the regression fit data
provided or, easily refit and store the regression model via
\code{stats::update}.
\begin{Schunk}
\begin{Sinput}
R> preferable_cp4 <- cpr_run4[[4]]
R> str(preferable_cp4)
\end{Sinput}
\begin{Soutput}
List of 12
$ cp :Classes 'tbl_df', 'tbl' and 'data.frame': 7 obs. of 2 variables:
..$ xi_star: num [1:7] -1 -0.6889 -0.3703 0.0298 0.3854 ...
..$ theta : num [1:7] -0.0477 -0.4992 -0.5588 -0.1586 0.9388 ...
$ xi : num [1:11] -1 -1 -1 -1 -0.0668 ...
$ iknots : num [1:3] -0.0668 -0.0443 0.2006
$ bknots : num [1:2] -1 1
$ order : num 4
$ call : language cp(formula = log10(pdg) ~ bsplines(day, iknots =
   c(-0.0667568176695966, -0.0442920251104394, 0.200576701268743)) + (1 | id), data
   = spdg, method = lmer, check_rank = FALSE)
$ keep_fit : logi FALSE
$ fit : logi NA
$ loglik : num 8523
$ rmse : num 0.155
$ coefficients: num [1:7] -0.0477 -0.4992 -0.5588 -0.1586 0.9388 ...
$ vcov : num [1:7, 1:7] 1.23e-04 8.09e-05 1.15e-04 1.01e-04 1.08e-04 ...
- attr(*, "class")= chr [1:2] "cpr_cp" "list"
\end{Soutput}
\begin{Sinput}
R> class(preferable_cp4$fit)
\end{Sinput}
\begin{Soutput}
[1] "logical"
\end{Soutput}
\begin{Sinput}
R> preferable_cp4 <- update(preferable_cp4, keep_fit = TRUE)
R> class(preferable_cp4$fit)
\end{Sinput}
\begin{Soutput}
[1] "lmerMod"
attr(,"package")
[1] "lme4"
\end{Soutput}
\end{Schunk}

\subsection{Exploring Additional Model Spaces}
Because CPR is a relatively fast method for model selection we can explore other
polynomial orders and/or knot sequences.  \code{cpr::update_bsplines} will allow
the end user to quickly modify the \code{cpr::bsplines} call within the
\code{formula} element of a \code{cpr::cp} call.  In the following, we set up
\code{initial_cp3} and \code{initial_cp2} for third and second order splines.
The modified \code{df} are such that the knot sequences between
\code{initial_cp4}, \code{initial_cp3}, and \code{initial_cp2} are the same.
The updated initial control polygons are used to seed to additional CPR runs.
\begin{Schunk}
\begin{Sinput}
R> initial_cp3 <- update_bsplines(initial_cp4, df = 53, order = 3)
R> initial_cp2 <- update_bsplines(initial_cp4, df = 52, order = 2) 
R> cpr_run3 <- cpr(initial_cp3)
R> cpr_run2 <- cpr(initial_cp2)
\end{Sinput}
\end{Schunk}

\begin{figure}
  \centering
\begin{Schunk}

\includegraphics[width=\maxwidth]{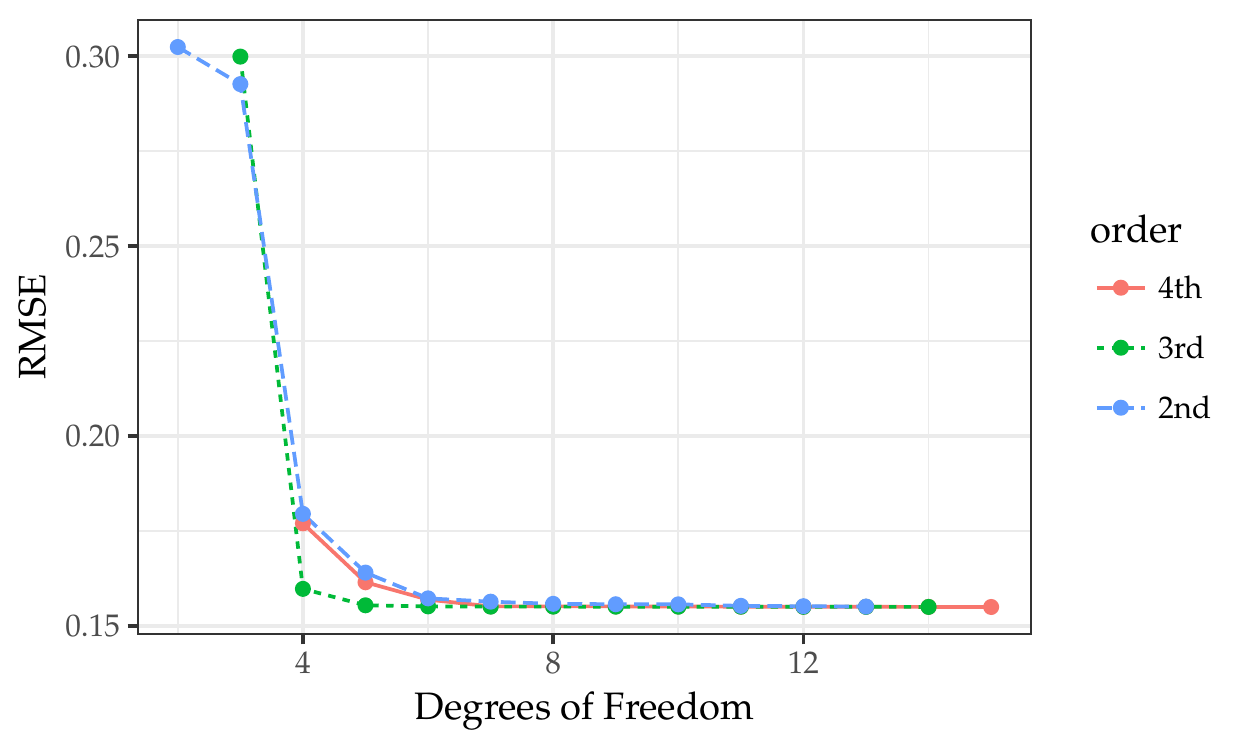} \end{Schunk}
\caption{RMSE by degrees of freedom (polynomial order plus number of internal
knots) for three CPR runs. The preferable model appears to be a 3rd order
B-spline with five degrees of freedom (two internal knots) as it has the lowest
RMSE for the degree of freedom, and there is no noticable decrease in RMSE for
additional degrees of freedom.}
\label{fig:other_cpr_run_summary}
\end{figure}

To select a preferable model we will compare the RMSE by degrees of freedom as
in Figure~\ref{fig:other_cpr_run_summary}.
\begin{Schunk}
\begin{Sinput}
R> list(cpr_run4, cpr_run3, cpr_run2) %>%
+    lapply(summary) %>%
+    bind_rows(.id = "order") %>%
+    mutate(order = factor(order, 1:3, c("4th", "3rd", "2nd"))) %>%
+    filter(index < 13) %>% 
+    ggplot() +
+    theme_bw() +
+    aes(x = dfs, y = rmse, color = order, linetype = order) +
+    geom_path() +
+    geom_point() +
+    ylab("RMSE") +
+    xlab("Degrees of Freedom")
\end{Sinput}
\end{Schunk}

The third order B-spline with five degrees of freedom, that is two internal
knots, is the preferable model as it has the lowest RMSE and degrees of freedom.

\begin{Schunk}
\begin{Sinput}
R> # cpr_run3 with five degrees of freedom: 3rd order + 2 knots = 5 df.  Model
R> # index 3 has two knots.
R> preferable_model <- cpr_run3[[3]]
\end{Sinput}
\end{Schunk}

\section{Control Net Reduction} \label{sec:cnr}
Control net reduction, CNR, is the natural extension from uni-variable B-splines
to multi-variable B-splines.   This section is a brief overview of the extension
and tools in the \pkg{cpr} package for use in this case.

\subsection{Multi-variable B-Splines}
We generalize \eqref{eq:cpr_general_model} to have a multi-variable function as
the varying mean element, \ie,
\begin{equation}
  \label{eq:cnr_general_model}
  \bs{y} = f \left( \bs{x}_1, \bs{x}_2, \ldots, \bs{x}_m \right) + \bs{Z}_{f} \bs{\beta} + \bs{Z}_{r} \bs{b} +
  \bs{\epsilon}.
\end{equation}

Multi-variable B-spline functions are constructed by tensor products of
uni-variable B-splines bases, henceforth referred to as marginal B-splines.  We
define the multi-variable B-spline function in terms of matrix arithmetic and as
an algebraic formula, the latter will be useful in assessing the influence of a
knot on a tensor product.

We denote a multi-variable $m$-dimensional B-spline function, built on $m$ B-spline bases
$\bs{B}_{k_1, \bs{\xi}_1} \left(\bs{x}_1\right),$
$\bs{B}_{k_2, \bs{\xi}_2} \left(\bs{x}_2\right),$ \ldots,
$\bs{B}_{k_m, \bs{\xi}_m} \left(\bs{x}_m\right),$ as
\begin{equation}
  \label{eq:bspline_multivariable_function}
  f\left( \bs{X} \right) = \mathscr{B}_{\bs{K}, \bs{\Xi}} \left(\bs{X}\right)
  \bs{\theta}_{\bs{\Xi}},
\end{equation}
where 
$\bs{K} = \left\{k_1, k_2, \ldots, k_m \right\}$, denotes the set of polynomial
orders,
$\bs{\Xi} = \left\{\bs{\xi}_1, \bs{\xi}_2, \ldots, \bs{\xi}_m \right\},$ is the
set of knot sequences, 
$\bs{\theta}_{\bs{\Xi}}$ is a $\prod_{i=1}^{m}
\left(\card{\bs{\xi}_i} - k_i \right) \times 1$ column vector of
regression coefficients,  and
$\bs{X}$ is the observed data
\begin{equation}
  \bs{X} = \begin{pmatrix}
    x_{11} & x_{21} & \cdots & x_{m1} \\ 
    x_{12} & x_{22} & \cdots & x_{m2} \\ 
    \vdots & \vdots & \ddots & \vdots \\
    x_{12} & x_{22} & \cdots & x_{mn}
  \end{pmatrix}.
\end{equation}

The basis for multi-variable B-splines is constructed by a recursive algorithm.
The base case for $m = 2$ is
\begin{equation}
  \label{eq:tensor_product_base_case}
  \mathscr{B}_{ \left\{k_1, k_2\right\}, \left\{ \bs{\xi}_1, \bs{\xi}_2 \right\}} \left( \bs{x}_1, \bs{x}_2 \right) = 
  \left(\bs{1}^T_{\card{\bs{\xi}_2} - k_2} \otimes \bs{B}_{k_1, \bs{\xi}_1} \left(\bs{x}_1 \right)  \right) \odot
  \left(\bs{B}_{k_2, \bs{\xi}_2} \left( \bs{x}_2 \right) \otimes
  \bs{1}^{T}_{\card{\bs{\xi}_1} - k_1} \right),
\end{equation} 
where $\odot$ is the element-wise product, $\otimes$ is a
Kronecker product, and $\bs{1}_n$ is a $n \times 1$ column vector of 1s.  The
two Kronecker products define the correct dimensions for the entry-wise product.
The tensor product matrix as the same number of rows as the two input matrices
and the columns are generated by all the pairwise products of the columns of the
two input matrices.
The general case for $m > 2,$ the matrix $\mathscr{B}_{\bs{K}, \bs{\Xi}} \left( \bs{X} \right)$ is defined by
\begin{equation}
  \label{eq:tensor_product_recursive}
    \mathscr{B}_{\bs{K}, \bs{\Xi}} \left( \bs{X} \right) =
    \left( \bs{1}_{\card{\bs{\xi}_m} - k_m}^{T} \otimes \mathscr{B}_{\bs{K} \backslash
  k_{m}, \bs{\Xi} \backslash \bs{\xi}_m} \left(\bs{X} \backslash \bs{x}_{m} \right) \right)
    \odot
    \left( \bs{B}_{k_m, \bs{\xi}_m}\left(\bs{x}_m\right) \otimes
    \bs{1}_{\prod_{i = 1}^{m-1} \left(\card{\bs{\xi}_i} - k_i \right)}^{T}
  \right).
\end{equation}

It is possible to write \eqref{eq:bspline_multivariable_function} as a set of
summations as follows:
\begin{equation}
  \label{eq:bspline_multivariable_function_sum}
  \begin{split}
    f\left( \bs{X} \right) & = \mathscr{B}_{\bs{K}, \bs{\Xi}} \left( \bs{X} \right) \\
   & =
    \sum_{j_1 = 1}^{\card{\bs{\xi}_1} - k_1}
    \sum_{j_2 = 1}^{\card{\bs{\xi}_2} - k_2}
  \cdots
  \sum_{j_m = 1}^{\card{\bs{\xi}_m} - k_m}
  \bs{B}_{j_1, k_1, \bs{\xi}_1} \left( \bs{x}_1 \right)
  \bs{B}_{j_2, k_2, \bs{\xi}_2} \left( \bs{x}_2 \right)
  \cdots
  \bs{B}_{j_m, k_m, \bs{\xi}_m} \left( \bs{x}_m \right)
  \theta_{\bs{\Xi}, j_1, j_2, \ldots, j_m} \\
  & =
    \sum_{j_1 = 1}^{\card{\bs{\xi}_1} - k_1} \bs{B}_{j_1, k_1, \bs{\xi}_1} \left( \bs{x}_1 \right) 
  \underbrace{
    \sum_{j_2 = 1}^{\card{\bs{\xi}_2} - k_2}
  \cdots
  \sum_{j_m = 1}^{\card{\bs{\xi}_m} - k_m}
  \bs{B}_{j_2, k_2, \bs{\xi}_2} \left( \bs{x}_2 \right)
  \cdots
  \bs{B}_{j_m, k_m, \bs{\xi}_m} \left( \bs{x}_m \right)
  \theta_{\bs{\Xi}, j_1, j_2, \ldots, j_m}
  }_{\text{polynomial coefficients}} \\
  & = 
  \text{diag}\left(
    \bs{B}_{k_1, \bs{\xi}_1}\left(\bs{x}_1\right)\bs{\theta}_{\bs{\Xi}\backslash \bs{\xi}_1} \left( \bs{X} \backslash \bs{x}_1
    \right)
  \right).
  \end{split}
\end{equation}
When the data input is a single observation, that is a tuple $\left(x_1, x_2,
\ldots, x_m\right) \in \text{rows}\left(\bs{X}\right)$ the last line is a
singleton and the diag operation is redundant.  Equation
\eqref{eq:bspline_multivariable_function_sum} is critical in the extension from
the uni-variable control polygon reduction method to the multi-variable control
polygon reduction method.  By conditioning on $m-1$ marginals, the
multi-variable B-spline becomes a uni-variable B-spline in terms of the $m^{th}$
marginal.  Thus, the metrics and methods developed
for uni-variable B-splines can be applied to multi-variable B-splines.

To simplify the explanation consider a $m = 2$ dimensional B-spline.  The
concepts and notation extend to $m > 2$ with ease.
If we condition on $x_2,$ the two-dimensional B-spline simplifies to a
uni-variable spline in $x_1,$ \ie,
\begin{equation}
  \label{eq:conditional_uni_var_bspline}
  \mathscr{B}_{\bs{K}, \bs{\Xi}} \left(x_1 | x_2 \right) \bs{\theta}_{\bs{\Xi}} = 
  \bs{B}_{k_1, \bs{\xi}_1}\left(x_1\right)\bs{\theta}_{\bs{\Xi}\backslash \bs{\xi}_1}
    \left( x_2 \right).
\end{equation}
In terms of the control net, the conditioning on $x_2$ creates a slice across the
net and yields the control polygon $\CP_{k_1, \bs{\xi}_1, \bs{\theta}_{\bs{\Xi} \backslash \bs{\xi}_1}\left(x_2\right)}.$

The influence weight of $\xi_{1j} \in \bs{\xi}_1$
is the Euclidean distance between the ordinates $\bs{\theta}_{\bs{\Xi} \backslash \bs{\xi}_1}\left(x_2\right)$
and $\bs{\theta}_{\bs{\Xi} \backslash \left( \left( \bs{\xi}_1 \backslash \xi_{1j} \right) \cup \xi_{1j} \right)}\left(x_2\right).$
\begin{equation}
  \label{eq:cnr_conditional_influence_weights}
  w_{1j | x_2} =
  \left\lVert
    \left(
      \bs{I} - 
      \bs{W}_{k_1, \bs{\xi}_1}\left(\xi_{1j}\right)
      \left( \bs{W}_{k_1, \bs{\xi}_1}^{T}\left(\xi_{1j}\right) \bs{W}_{k_1, \bs{\xi}_1}\left(\xi_{1j}\right) \right)^{-1} \bs{W}_{k_1, \bs{\xi}_1}^{T}\left(\xi_{1j}\right)
    \right)
    \bs{\theta}_{\bs{\Xi} \backslash \bs{\xi}_1} \left( x_2 \right)
  \right\rVert_2.
\end{equation}

The conditional influence weight, \eqref{eq:cnr_conditional_influence_weights},
is used to get the influence weight of $\xi_{1j}$ on
$\CP_{k_1, \bs{\xi}_1, \bs{\theta}_{\bs{\Xi} \backslash
\bs{\xi}_1}\left(x_2\right)}.$  That is, the relative influence weight of
$\xi_{1j}$ is the maximum influence weight over a set of $p$ values for $x_2.$
\begin{equation}
  \label{eq:cnr_influence_weight}
  w_{1j} = \max_{x_2 \in \bs{U}} w_{1j | x_2},
\end{equation}
where
\begin{equation}
  \bs{U} = \left\{u : \min \left( x_2 \right) + \frac{ \left\{1, 2, \ldots, p \right\} }{p+1} \left( \max\left( x_2  \right) - \min\left(x_2\right) \right) \right\}.
\end{equation}
We recommend and have set the default in \pkg{cpr} to use $p = 20$ values for each marginal.

\subsection{The CNR Algorithm}
\begin{enumerate}
  \item Define the initial $\bs{K}$ and $\bs{\Xi}$ set for
    the initial tensor product.
  \item Use an appropriate regression modeling approach to fit a regression
    model for~\eqref{eq:cnr_general_model}. 
  \item \label{cnr_step:build} Construct the control net for the current
    set of knots sequences and regression coefficients.
  \item Use \eqref{eq:cnr_influence_weight} to find the influence weight
    for all internal knots on the marginals you are interested in reducing.
  \item Coarsen the knot sequence by removing the knot with
    the smallest influence weight.
  \item\label{cnr_step:refit} Refit the regression model using the coarsened knot
    sequences.
  \item Repeat steps \ref{cnr_step:build} through \ref{cnr_step:refit} until all
    internal knots have been removed.
  \item\label{cnr_step:selection} Select the preferable model by visual inspection
    of diagnostic graphics.
\end{enumerate}

CNR is applied to all margins of interest at once.  The knot with the lowest
influence weight is omitted at each step, regardless of the margin it originates
from.

\subsection{Implementation Issue: Dimensionality}
The CNR algorithm is not immune to the curse of dimensionality.  The total
degrees of freedom consumed by the tensor product is the product of the degrees
of freedom of each marginal B-spline.  This fact needs to be considered when
starting the search for a parsimonious model.  Consider a two-dimensional double
cubic B-spline with fifty internal knots on each marginal.  This construction
generates a regression model with $54 \times 54 = 2916$ regression
coefficients.

This is important because a common method for solving regression problems is to
use a QR decomposition of the design matrix.  QR decomposition is an $O(n^3)$
algorithm where $n$ is the number of regression coefficients.  It will take a
considerable amount of computational resources to estimate the $2916$
regression parameters, and may not be feasible for some computing environments.

The size of the initial model is even more problematic for three-dimensional
B-splines.  Setting up a search for a set of parsimonious knot sequences with
three cubic B-splines, each with initially fifty internal knots would require
$157,464$ regression coefficients.  This is unreasonable.

We suggest that for $m \geq 3$ to model a primary explanatory variable via
uni-variable methods such as control polygon reduction, and then use the
resulting uni-variable B-spline as a static marginal in the multi-variable
B-spline.  Example follows in the next section.

\subsection{Use In The cpr Package}
The use of the CNR algorithm in the \pkg{cpr} package is very similar to the use
of the CPR algorithm.  Use \code{cpr::btensor} to generate the multi-variable
B-spline, \code{cn} to generate the control net, \code{cnr} to apply the CNR
algorithm, and \code{plot} to see diagnostic plots.  A simple example, fitting a
two-dimensional B-spline over \code{day} and \code{age}, with two fourth order
splines is below.  The example also shows an updated version of the initial
control net with a third order B-spline for \code{day} and fourth order spline
for \code{age}.
\begin{Schunk}
\begin{Sinput}
R> initial_cn44 <-
+    cn(log10(pdg) ~ btensor(list(day, age), df = list(24, 24)) + (1 | id),
+       data = spdg, method = lmer) 
R> initial_cn34 <- update_btensor(df = list(23, 24), order = list(3, 4))
R> cnr_run44 <- cnr(initial_cn44)
R> cnr_run34 <- cnr(initial_cn34)
R> plot(cnr_run44)
\end{Sinput}
\end{Schunk}

As noted in the prior subsection, building $m$-dimensional B-splines, $m \geq
3,$ can be difficult simply due to the number of degrees of freedom required to
build the basis.  For the \code{spdg} data, it might be reasonable to find a
parsimonious B-spline for PDG by \code{day} via CPR, and then use that result as
a foundation for the higher dimensional B-spline and CNR run.  For example,
build the initial knot sequences for \code{day}, \code{age}, \code{ttm} B-spline
based on the preferable model from the prior section.

\begin{Schunk}
\begin{Sinput}
R> init_iknots <-
+    list(cpr_run3[[3]]$iknots,
+         trimmed_quantile(spdg$age, prob = 1:10/11),
+         trimmed_quantile(spdg$ttm, prob = 1:10/11))
\end{Sinput}
\end{Schunk}

Construct the initial control net, and then run CNR on only the \code{age} and
\code{ttm} margins and the \code{day} margin has already been specified.
\begin{Schunk}
\begin{Sinput}
R> init_cn <- cn(log10(pdg) ~ btensor(list(day, age, ttm),
+                                     iknots = init_iknots,
+                                     order  = list(3, 2, 2)) + (1 | id),
+                data = spdg,
+                method = lmer)
R> cnr_run <- cnr(init_cn, margin = 2:3) 
\end{Sinput}
\end{Schunk}

\section{Discussion}\label{sec:discussion}
The \pkg{cpr} package provides a clean and user friendly interface for
implementing our Control Polygon Reduction and Control Net Reduction algorithms
for B-spline regression model selection.

The CPR and CNR algorithms take a novel approach to regression model selection.
Instead of focusing on the likelihood function we focus on the geometry of the
control polygons and control nets to determine the influence of any one
particular knot.  That said, the CPR algorithm has been shown to be able to
select models with fewer degrees of freedom and with superior fit statistics
compared to a forward step model selection approach.
The model space
searched by CPR is larger than the model space searched by a forward step
model selection approach. The additional calculations for determining knot
influence require few computation resources and are done quickly.  Thus, CPR
requires negligibly more time to run than a forward step selection approach. 

Compared to a comparable likelihood based backward-step selection approach, CPR
picks models which are on average equivalent to the likelihood based selected
models on a degree of freedom  for degree of freedom basis.  The time required
to run a likelihood based backward-step selection approach can be considerable,
CPR is much faster.

Overall, CPR can quickly select a parsimonious model with preferable fit
statistics to likelihood based model selection approaches.  Further, as was
demonstrated in this manuscript, since CPR requires very little time to run,
exploration of B-splines or different polynomial orders is feasible.  Fourth
order B-splines are commonly used as they have twice differentiable and
relatively smooth.  However, while perhaps not as `smooth', lower order splines
may fit the data more efficiently than higher order splines.

CNR provides analysts many of the same benefits as CPR, fast model selection
from a very large model space.  The tools allow for exploration of complex
non-linear multi-variable functions.

Continued development of
the package can be tracked and contributed to at
\url{https://github.com/dewittpe/cpr}.  The package is also hosted on the
Comprehensive R Archive Network (CRAN) at
\url{https://cran.r-project.org/package=cpr}.

\section{Acknowledgements}
The development of the \pkg{cpr} package was part of Peter DeWitt's Ph.D.\
dissertation under the advisement of co-authors Drs. MaWhinney and Carlson.  We
thank Dr. Edward Bedrick for his help in framing the influence weight heuristic.
Partial financial support for this work was provided by NIH grants 1 R03
DA026743, Forster and MaWhinney; and 1 R01 DA030495, Forster and MaWhinney.

\bibliography{references}

\end{document}